\documentclass{llncs}
\usepackage{amssymb,amsmath}
\usepackage{graphicx} 
\usepackage{algorithm}
\usepackage{algorithmic}
\usepackage{booktabs}
\usepackage{subfigure}
\usepackage{float}
\usepackage{hyperref}
\usepackage{cleveref}
\usepackage{comment}
\usepackage{wrapfig}

\title{Approximating Heavy-Tailed Distributions with a Mixture of Bernstein Phase-Type and Hyperexponential Models}

\author{Abdelhakim Ziani\inst{1} \and Andr\'as Horv\'ath\inst{2} \and Paolo Ballarini\inst{1} }

\institute{
Université Paris Saclay, Lab. MICS, CentraleSupélec,  Gif-sur-Yvette, France 
\email{\{hakim.ziani,paolo.ballarini\}@centralesupelec.fr}
\and 
Università di Torino, Torino, Italy\\
\email{horvath@di.unito.it}
}

\date{}

\begin{document} 
\maketitle
\begin{abstract}
Heavy-tailed distributions, prevalent in a lot of real-world applications such as finance, telecommunications, queuing theory, and natural language processing, are challenging to model accurately owing to their slow tail decay. Bernstein phase-type (BPH) distributions, through their analytical tractability and good approximations in the non-tail region, can present a good solution, but they suffer from an inability to reproduce these heavy-tailed behaviors exactly, thus leading to inadequate performance in important tail areas. On the contrary, while highly adaptable to heavy-tailed distributions, hyperexponential (HE) models struggle in the body part of the distribution. Additionally, they are highly sensitive to initial parameter selection, significantly affecting their precision.

To solve these issues, we propose a novel hybrid model of BPH and HE distributions, borrowing the most desirable features from each for enhanced approximation quality. Specifically, we leverage an optimization to set initial parameters for the HE component, significantly enhancing its robustness and reducing the possibility that the associated procedure results in an invalid HE model. Experimental validation demonstrates that the novel hybrid approach is more performant than individual application of BPH or HE models. More precisely, it can capture both the body and the tail of heavy-tailed distributions, with a considerable enhancement in matching parameters such as mean and coefficient of variation. Additional validation through experiments utilizing queuing theory proves the practical usefulness, accuracy, and precision of our hybrid approach.
\end{abstract}

\section{Introduction}
Heavy-tailed distributions are common in real-world domains like finance, tele\-communications, queuing theory, and NLP, where rare or extreme events significantly influence system behavior \cite{foss2011introduction}. Markov models, while advantageous due to their analytical and numerical tractability \cite{stewart1994markov}, struggle to represent these distributions accurately in the tail region.

Bernstein Phase-Type (BPH) distributions \cite{HorvathV23,HORVATH2025102480}, a class of distributions defined in terms of Markov chains, offer fast parameter estimation and precise approximation of the body of a distribution, but fall short in capturing heavy tails. Hyperexponential (HE) distributions, on the other hand, can better approximate slowly decaying tails when used with proper parameters \cite{[FELD98a]} but (i) they are highly sensitive to initialization, which can hinder performance, and (ii) they form an inflexible family of distributions (their probability density function can only be monotone decreasing). 

To address these limitations, we propose a hybrid model combining BPH and HE distributions. This approach leverages the strengths of both: the approximation efficiency of BPH distributions and the tail-fitting flexibility of HE distributions. To improve the robustness of the HE component, we employ the Adam optimizer to enhance parameter initialization. This paper presents the hybrid BPH\_HE model and the optimization method, evaluates performance, and validates results through queuing theory applications. 

A similar approach was presented in \cite{HoTe00}, where the authors proposed combining acyclic phase-type (APH) distributions with HE distributions. The improvements over that work are as follows: (i) instead of APH, we apply BPH distributions, which allow for much faster and more precise approximation of the body, and (ii) we automate the parameter initialization of the HE distributions that capture the tail, whereas in \cite{[FELD98a]} and \cite{HoTe00}, this process is left to a non-trivial manual setting.

The remainder of the paper is organized as follows: 
\Cref{sec:related} reviews related work, 
\Cref{sec:back} gives background, 
\Cref{sec:meth} details the hybrid model and the optimization strategy, 
\Cref{sec:exp} covers experiments, 
\Cref{sec:queue} application results, and 
\Cref{sec:conc} concludes the study.

\section{Related Work}\label{sec:related}

Phase-type (PH) distributions have been introduced in \cite{[NEUT75]} to model non-ex\-po\-nen\-tial durations by combination of exponential distributions, maintaining thus the Markov property of the underlying stochastic process. The application of PH distributions requires fitting (also called matching) algorithms, that is, methods that, given a distribution, provide a PH distribution that closely resembles it. Such methods either aim
\begin{enumerate}
    \item to match statistical properties extracted from the distribution (typically moments) or
    \item to minimize a distance measure that considers the whole distribution.
\end{enumerate}

A seminal work among those belonging to the first category is \cite{johnson1989matching}, where the authors use a mixture of Erlang distributions to match the first three moments of a given distribution. This approach was refined in \cite{bobbio2005matching}, providing a method to match any valid first three moments with an acyclic PH distribution of minimal size. Matching of more than three moments was tackled in \cite{horvath2007matching}. While the above approaches match the moments exactly, there have been proposals to approximate matching of the moments based on optimization \cite{BuKr09,sherzer2025unconstrainedoptimizationapproachmoment}. A common difficulty of all these approaches is that a full characterization of the feasible region of moments of PH distributions is not available. The known results in this direction are gathered in \cite{horvath2024phase}.

Approaches belonging to the second category are most often based on the maximum likelihood principle, that is, they minimize the Kullback–Leibler divergence. This can be done either by using the Expectation-Maximization (EM) algorithm, like in \cite{asmussen1996fitting,okamura2011refined}, or by more direct maximization of the likelihood function, as done in \cite{[BOBB94a]}, where also a benchmark for the evaluation of fitting methods is proposed.

For what concerns fitting heavy tailed distributions, general approaches belonging to either the first or the second category fail to result in PH distributions with satisfactory behavior. For this reason, \cite{[FELD98a]} developed a heuristic approach, which we describe in detail in Sec. \ref{Section:he}, whose shortcoming is that it is applicable only to distributions with a monotone probability density function. To overcome this, in \cite{HoTe00} the authors proposed to combine the model proposed in \cite{[FELD98a]} with an acyclic PH distribution whose parameters are estimated based on the maximum likelihood principle.

Models constructed through the composition of PH distributions (see, e.g., \cite{Haddad1997}) result in Markov chains, which can be analyzed using established techniques developed for such models \cite{stewart1994markov}. In the area of queuing systems, a wide range of methods have been developed to analyze models that employ PH distributions. The seminal work in this regard is \cite{neuts1981matrix} while \cite{latouche1999introduction} provides a modern, comprehensive treatment. A first effort to propose a tool to analyze Petri net models with PH timing was described in \cite{Cu85}.

Among others, tools for PH fitting are \cite{Horvath2002PhFit,HyperStarTool} while \cite{Horvath2017BuTools} provides both moment matching algorithms and methods to analyze queues with PH timing.

\section{Background\label{sec:back}}

\noindent
{\bf Notations}. Given a positive continuous random variable, $X$, we denote by $f(x)$ its probability distribution function (PDF), $F(x)=\int_{0}^x f(x) dx $ its cumulative distribution function (CDF)  and $\bar{F}(x) = 1 - F(x)$ its complementary cumulative distribution function (CCDF).

\subsection{Heavy tailed distributions}

A heavy-tailed distribution is a probability distribution whose tail decays more slowly than that of an exponential distribution, implying that very large values can occur with non-negligible probability \cite{foss2011introduction}. Mathematically, a distribution \(F(x)\) is said to be heavy-tailed if its CCDF satisfies 
\begin{equation}
    \lim_{x \to \infty} e^{\lambda x} \bar{F}(x) = \infty, \quad \forall \lambda > 0
\end{equation}
meaning that the probability of large values does not decay at least exponentially fast, leading to distributions with a significant probability mass in the tail region. In most practical applications, heavy-tailed distributions are defined on positive values.

Heavy-tailed distributions appear in a wide range of real-world phenomena, often characterizing situations where extreme values or rare events have a non-negligible impact. Examples include queuing theory \cite{AFOLALU20212884,HarcholBalter2021}, where in many fields such as networking, telecommunication, or cloud computing, service time is modeled by a heavy-tailed distribution. In finance, risk modeling frequently exhibits heavy tails \cite{riskfinanceht}, meaning extreme outcomes or losses occur more often than in a Gaussian model (where rare events happen with a probability near zero), which is crucial in portfolio management. In natural language processing (NLP), word frequency distributions in all spoken languages follow a heavy-tailed distribution, characterized by Zipf’s law \cite{Piantadosi2014}, where 20\% of the most frequent words make 80\% of the spoken or written language. This property is particularly important in large language models (LLMs), as they must balance learning common words and patterns while still capturing rare but meaningful linguistic structures \cite{llmzifs}. 


\subsection{Phase type distributions}

\begin{wrapfigure}[13]{R}{0.3\textwidth}
\vspace{-1cm}
  \begin{center}
\includegraphics[width=0.26\textwidth]{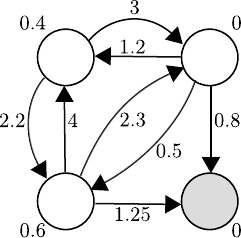}
  \end{center}
  \caption{A degree 3 PH distribution\label{fig:ph}}
\end{wrapfigure}
A phase-type (PH) distribution is a probability distribution resulting from the combination of exponential distributions. More formally, as introduced in \cite{[NEUT75]}, a degree $n$ continues PH distribution is given by the time to absorption in a continuous-time Markov chain (CTMC) having $n$ transient states (called phases) and one absorbing state. The most widely used representation of a PH distribution is a vector-matrix pair $(a,A)$ where the initial probability vector $a$ and the infinitesimal generator matrix $A$ only account for the transient states of the Markov chain\footnote{ This is because the intensities from the transient states to the absorbing one can be determined based on $A$ and we assume that the initial probability associated with the absorbing state is 0.}. A graphical example of a PH distribution is provided in Figure~\ref{fig:ph} with vector-matrix pair representation  
\begin{align}\nonumber
a=
  \begin{bmatrix}0.4 & 0 & 0.6
  \end{bmatrix},~~
  A=
  \begin{bmatrix}
    -5.2 & 3 & 2.2  \\
    1.2 & -2.5 & 0.5  \\
    4 & 2.3 & -7.55  \\
  \end{bmatrix}\,.
\end{align}

Given $a$ and $A$ of a PH distribution, the associated PDF, CDF, and CCDF are given by 
\begin{equation}\label{eq:phbasics}
f(x)=ae^{xA}(-A\mathbf{1}),~
F(x)=1-ae^{xA}\mathbf{1},\text{ and }
\bar{F}(x)=ae^{xA}\mathbf{1}
\end{equation}
where $\mathbf{1}$ denotes a column vector in which all entries are equal to 1.

Phase-type distributions form a dense family of distributions, meaning that any distribution can be approximated arbitrarily well by a PH distribution. However, since PH distributions are inherently light-tailed (i.e., their tails decay exponentially), classical parameter estimation techniques, such as maximum likelihood estimation or moment matching, typically fail to approximate heavy-tailed distributions accurately. To address this, PH distributions with special structures must be employed. From an application standpoint, PH distributions offer both analytical and numerical tractability, providing closed-form expressions for key metrics such as moments, tail probabilities, and Laplace transforms.


\subsection{Bernstein phase type distributions\label{sec:bph}}

Given a function $g: [0,1]\to\mathbb{R}$, its order-$n$ Bernstein polynomial (BP) approximation is given by
\begin{equation}\label{BPapprox}
BP_n(g;x):=\sum_{i=0}^n g\!\left(\frac{i}{n}\right)\cdot\binom{n}{i}x^i (1-x)^{n-i}\ ,
\end{equation}
which is a degree-$n$ polynomial.

\begin{figure}[b]
    \centering
        \includegraphics[width=0.8\textwidth]{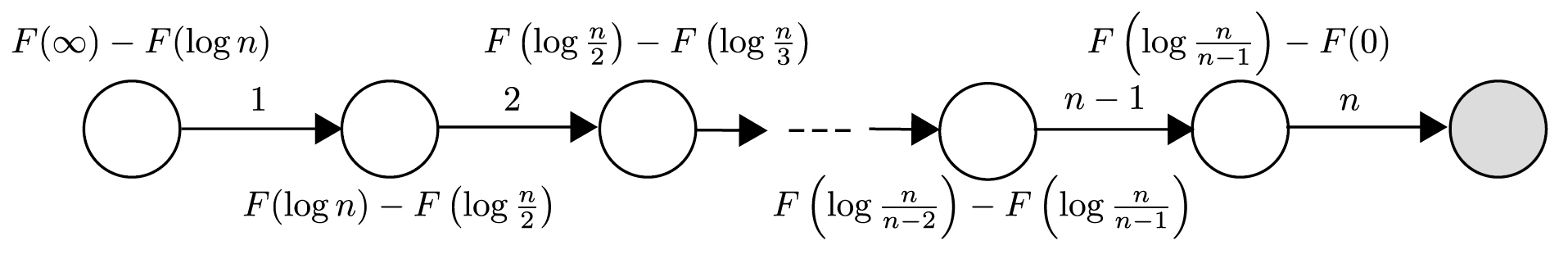} 
    \caption{Bernstein PH approximation of a CDF $F(x)$.}
    \label{fig:be_as_ph}
\end{figure}

 With the change of variable $y=e^{-x}$ (i.e., $x=-\log(y)$), which maps the interval $[0,1]$ onto $[0,\infty)$, one can approximate a function $f: [0,\infty)\to\mathbb{R}$. This leads to the Bernstein exponential (BE) approximations in the form of 
\begin{equation}\label{BEapprox}
BE_n(f;x):=\sum_{i=0}^n f\!\left(\log\left(\frac{n}{i}\right)\right) \cdot  \binom{n}{i} e^{-i x} (1-e^{-x})^{n-i}
\end{equation}
where the division by zero in case of $i=0$ is resolved by considering the limiting value of $f(x)$ as $x$ tends to infinity, that is, $f \left( \log \frac{n}{0} \right)=\lim_{x \rightarrow \infty}f(x)$. The approximation given in \cref{BEapprox} can be applied either to the PDF, to the CDF or to the CCDF of a probability distribution. In all cases, the approximation belongs to a class of PH distributions that we call Bernstein phase-type (BPH) distributions \cite{HORVATH2025102480}. In this paper, we apply CDF based approximation as it is somewhat more straightforward than using the PDF \cite{HORVATH2025102480}. \Cref{fig:be_as_ph} provides a graphical representation of the obtained PH distribution when \cref{BEapprox} is applied to the CDF $F(x)$. The vector of initial probabilities 
\[
a=\left[F(\infty)\!-\!F(\log n)~~~F(\log n)\!-\!F(\log (n/2))~~...~~F(\log (n/(n-1)))\!-\!F(0)\right]
\] 
are calculated directly based on the CDF while the intensities are fixed, in the sense that they do not depend on the CDF we approximate. If the CDF corresponds to a normalized distribution without mass at 0, then $F(\infty)=1$ and $F(0)=0$. In \cite{HORVATH2025102480} it was shown that applying \cref{BEapprox} to a CDF $F(x)$ or to its corresponding CCDF $\bar{F}(x)=1-F(x)$ results in the very same BPH distribution. The initial probabilities based on the CCDF are simply
\[
a=\left[\bar{F}(\log n)\!-\!\bar{F}(\infty)~~~\bar{F}(\log (n/2))\!-\!\bar{F}(\log n)~~...~~\bar{F}(0)\!-\!\bar{F}(\log (n/(n-1)))\right]
\]

\begin{figure}[t]
    \centering
    \begin{tabular}{cc}
        \includegraphics[width=0.454\textwidth]{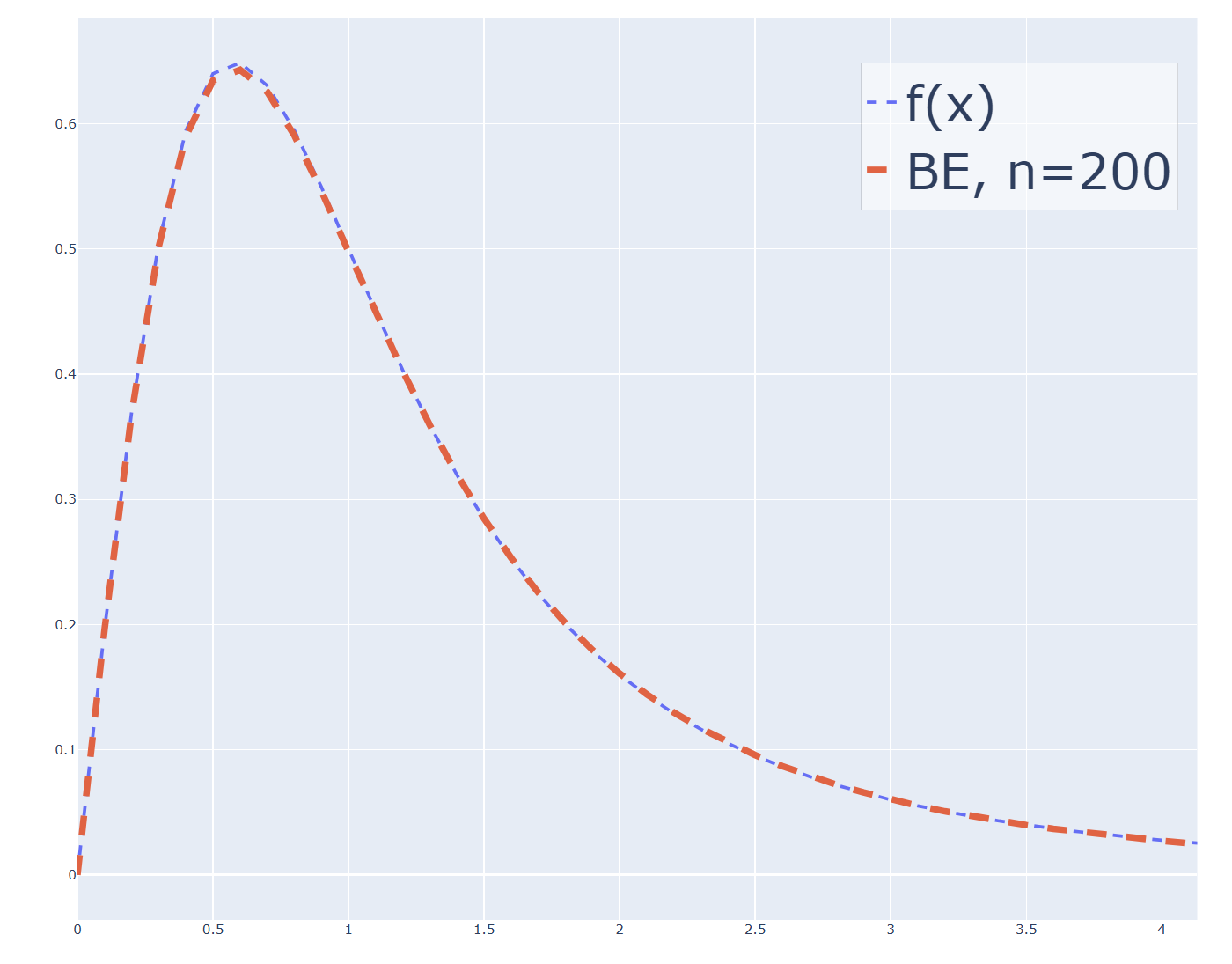} & 
        \includegraphics[width=0.45\textwidth]{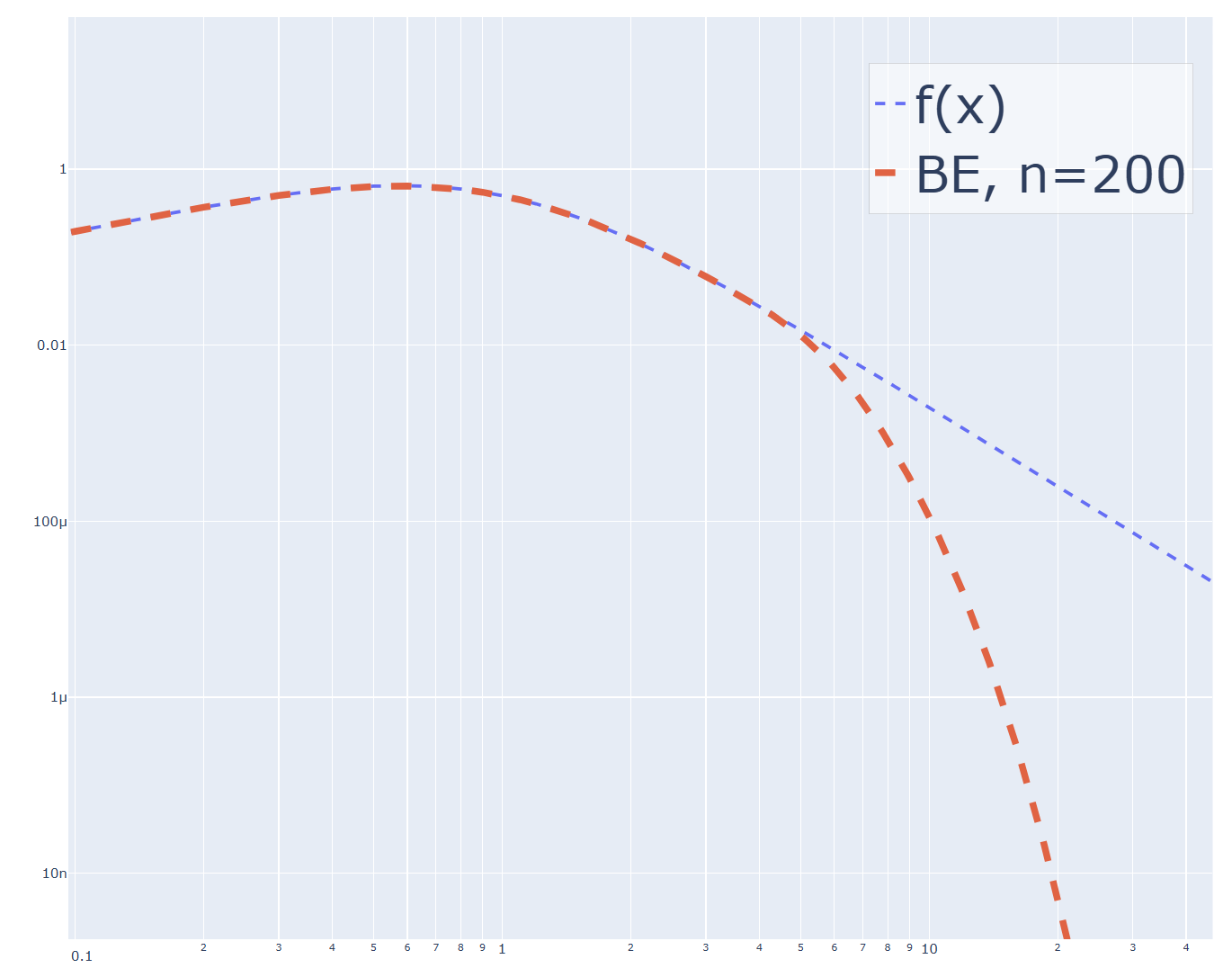} \\
        (a) body, linear plot & (b) tail, log scale plot
    \end{tabular}
    \caption{PDF resulting from a CDF-based BPH approximation of the Burr distribution with parameters $c=2, d=1$.}
    \label{fig:bph_burr}
\end{figure}

\Cref{fig:bph_burr} shows the PDF resulting from a CDF-based BPH approximation\footnote{We prefer to show the PDF because it highlights more the differences than plotting the CDF or the CCDF.} of the Burr distribution, whose PDF is
\[
f(x)=c\cdot d\cdot \frac{x^{c-1}}{(1+x^c)^{d+1}}~,
\]
using parameters $c=2, d=1$. A good approximation can be obtained in the body part of the distribution, especially when using a considerably large $n$ value. However, BPH models result in poor approximation in the tail region due to the fact that they lack the flexibility to capture the slow decay of heavy-tailed distributions.

\subsection{Hyperexponential approximation of heavy-tailed distributions}\label{Section:he}

In \cite{[FELD98a]} a method is proposed to approximate heavy-tailed distributions with a mixture of exponential distributions whose CCDF is in the form:
\begin{equation}\label{eq:he}
\bar{F}(x)=\sum_{i=1}^k p_i e^{-\lambda_i x}~~\mbox{with}~~
\sum_{i=1}^k p_i=1~~\mbox{and}~~0\leq p_i \leq 1, \lambda_i>0, i=1,2,...,k
\end{equation}
which is also a special case of the PH distributions, called the hyperexponential (HE) distribution.

The mixture in \cref{eq:he} has $2k-1$ free parameters ($k$ intensities and $k-1$ initial probabilities). 
The method aims to find parameters for the mixture such that the corresponding CCDF is close to the CCDF we need to approximate at $2k-1$ points, denoted by $x_1,x_2,...,x_{2k-1}$. Assuming that we have $x_1>x_2>...>x_{2k-1}$ and denoting by $\bar{F}(x)$ the CCDF to fit, the method proceeds as follows. The first two parameters, $p_1$ and $\lambda_1$, are chosen in such a way that we have
\begin{equation}\label{eq:p1l1}
p_1 e^{-\lambda_1 x_1}=\bar{F}(x_1)~~\mbox{and}~~p_1 e^{-\lambda_1 x_2}=\bar{F}(x_2)
\end{equation}
The further pairs, $p_i$ and $\lambda_i$ with $2\leq i \leq k-1$, are determined based on 
\begin{equation}\label{eq:pili}
p_i e^{-\lambda_i x_{2i-1}}=\bar{F}(x_{2i-1})
-\sum_{j=1}^{i-1} p_j e^{-\lambda_j x_{2i-1}}
~~\mbox{and}~~p_i e^{-\lambda_i x_{2i}}=\bar{F}(x_{2i})
-\sum_{j=1}^{i-1} p_j e^{-\lambda_j x_{2i}}
\end{equation}
That is, the effect of the already determined terms is taken into account by decreasing $\bar{F}(x)$. Finally, $p_k$ is computed as
\[
p_k=1-\sum_{i=1}^{k-1} p_k
\]
and $\lambda_k$ based on the last point $x_{2k-1}$ in such a way that
\[
p_k e^{-\lambda_k x_{2k-1}}=\bar{F}(x_{2k-1})
-\sum_{j=1}^{k-1} p_j e^{-\lambda_j x_{2k-1}}
\]
All the above calculations correspond to simple explicit expressions to determine the parameters. The basic idea behind the procedure is that the term $p_i e^{-\lambda_i x}$, whose parameters are determined based on $\bar{F}(x_{2i-1})$ and $\bar{F}(x_{2i})$, have a negligible effect at the points $x_j$ with $j>2i$ because of the exponential decay of the term itself.

If the points $x_i$ are not chosen appropriately, the calculations proposed in \cite{[FELD98a]} may lead to a non-valid hyperexponential distribution (e.g., with negative initial probabilities) or poor approximation. Since \cite{[FELD98a]} does not provide an algorithm for selecting these points, applying the method remains challenging.

\begin{figure}[H]
    \centering
    \begin{tabular}{cc}
        \includegraphics[width=0.44\textwidth]{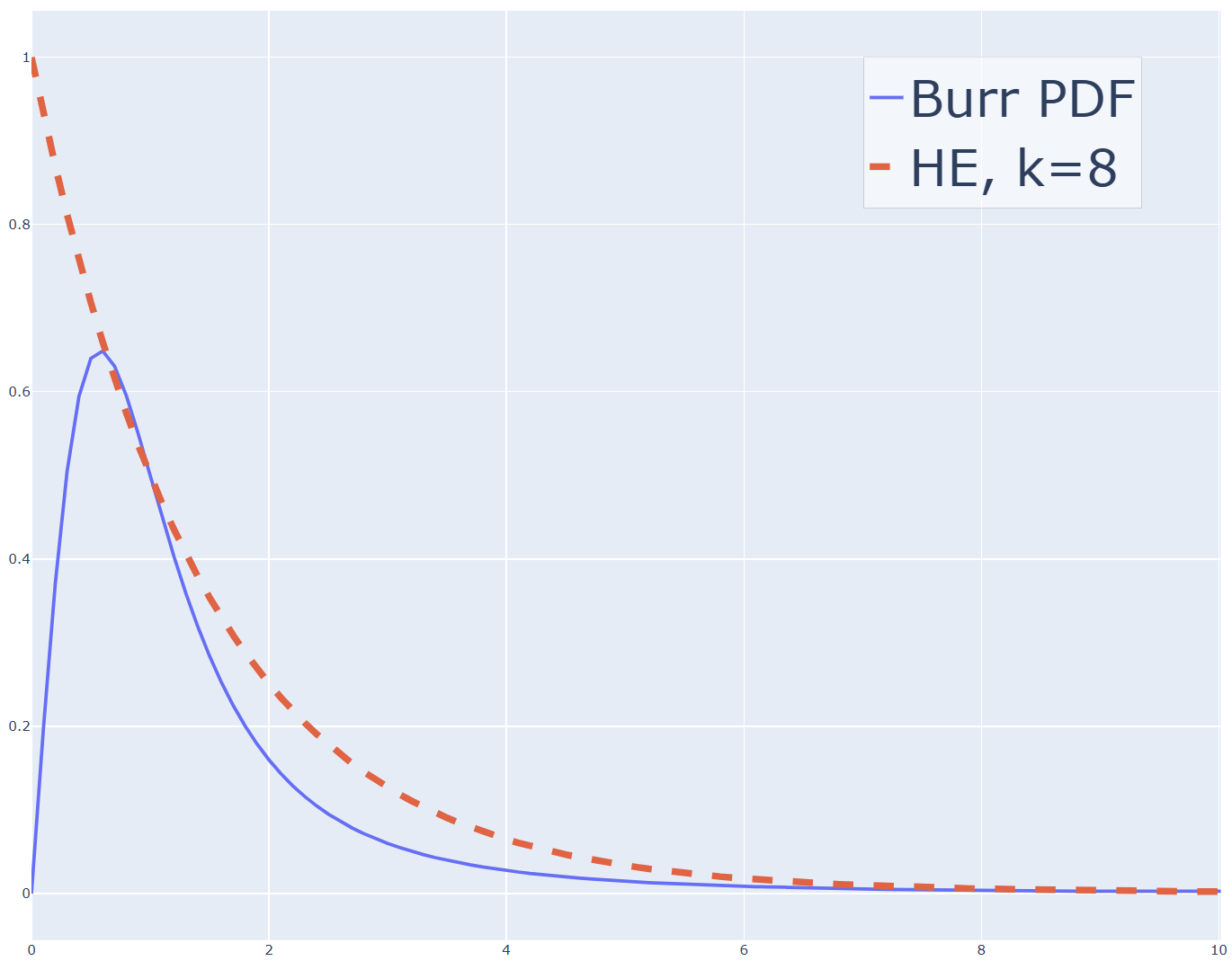} & 
        \includegraphics[width=0.445\textwidth]{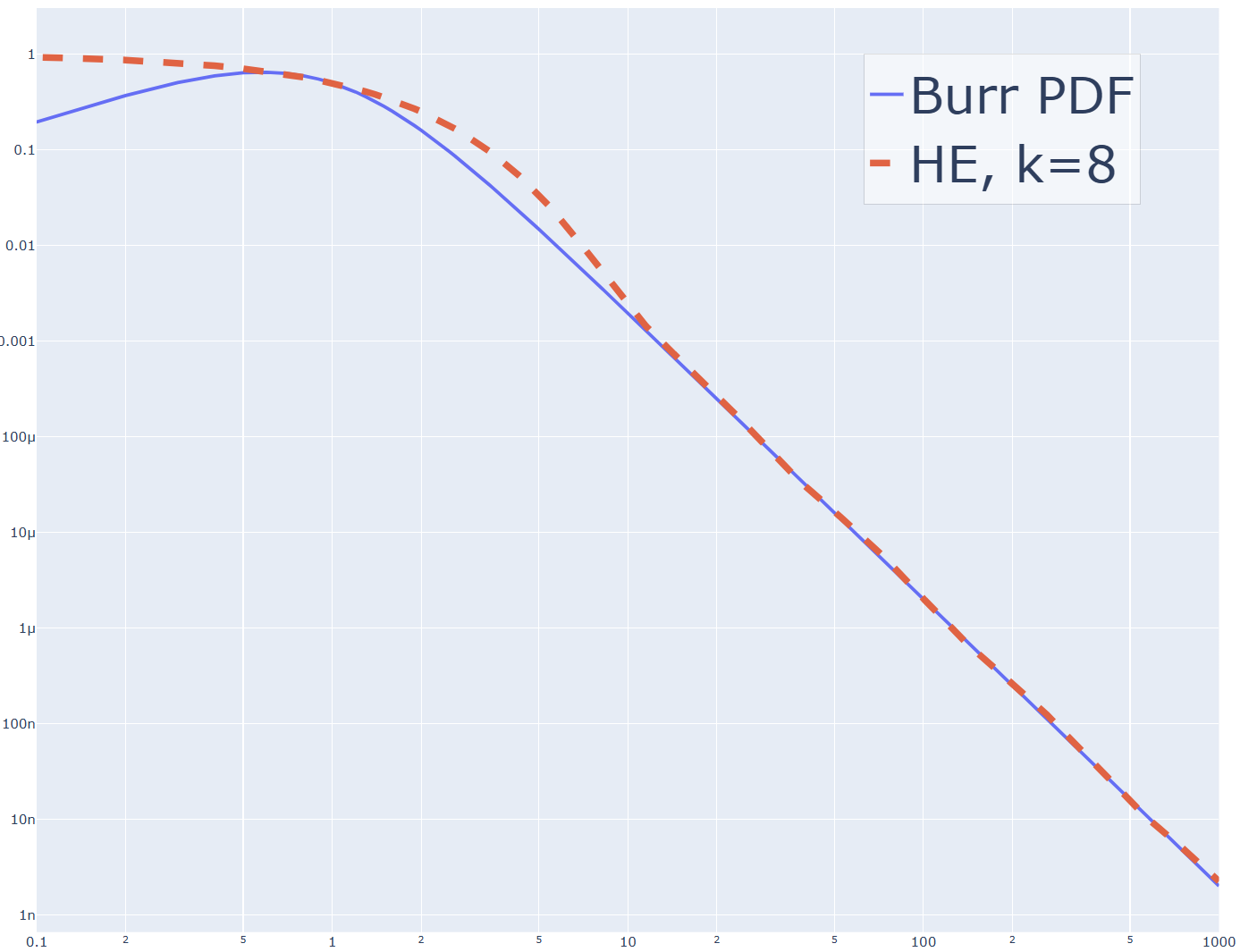} \\
        (a) body, linear plot & (b) tail, log scale plot
    \end{tabular}
\caption{PDF of the HE approximation of a Burr distribution}
    \label{fig:he_burr}
\end{figure}

\Cref{fig:he_burr} shows the HE approximation of the Burr distribution with $c=2,d=1$ using the method in \cite{[FELD98a]}. While the approximation captures the tail behavior well over a long interval, the figures reveal an important weakness of HE approximations: they come with a monotone PDF and hence cannot properly approximate the body region where the Burr PDF is non-monotonic.


\section{Methodology\label{sec:meth}}

\subsection{Approximation with mixture of BPH and HE distributions}

In order to overcome the limitations of the previously described models, we combine the HE approximation (\Cref{Section:he}) and the BPH approximation (\Cref{sec:bph}), exploiting the strength of the two methods. The resulting mixture, called BPH\_HE distribution, belongs to the class of PH distributions.

For what concerns the HE component that approximates the tail, the procedure is almost the same as the one described in \Cref{Section:he}. The only difference is that the sum of the initial probabilities associated with the exponential terms has to be less than 1, resulting in a defective HE distribution, which will be combined with an order $n$ BPH distribution to obtain the final BPH\_HE distribution. Accordingly, having one less constraint, the parameters $p_i,\lambda_i, i=1,...,k$ are determined based on $2k$ points instead of $2k-1$, and, instead of $\sum_{i=1}^k p_i=1$, we must have $\sum_{i=1}^k p_i<1$. As before, explicit expressions are used to calculate $p_i,\lambda_i, i=1,...,k$ that can be derived easily based on \cref{eq:p1l1} and \cref{eq:pili}. 

In order to combine the mixture of $k$ exponentials with a BPH, we introduce
\begin{equation}
    \bar{G}(x)=\bar{F}(x)-\sum_{i=1}^k p_i e^{-\lambda_i x}
\end{equation}
which is the CCDF $\bar{F}$ we aim to approximate minus the already determined contribution of the $k$ exponential terms. To obtain the BPH component of the mixture, we apply the approximation given in \cref{BEapprox} to $\bar{G}$. Overall, the mixture of the $k$ exponential terms (the HE component) and the order $n$ BPH component has the CCDF
\begin{equation}
\label{eq:FW_BPHccdf}
\begin{aligned}
\hat{\bar{F}}(x) ={} & \sum_{i=1}^{n} 
\bar{G}\left(\log \frac{n}{i}\right) 
\binom{n}{i} e^{-ix}(1 - e^{-x})^{n-i} 
+ \sum_{j=1}^k p_j e^{-\lambda_j x}
\end{aligned}
\end{equation}
where the first term is the contribution of the BPH part\footnote{With respect to \cref{BEapprox} the term with $i=0$ is not present since we approximate the function $\bar{G}$ which tends to 0.} while the second is that of the HE component.

The above CCDF corresponds to an order $n+k$ PH distribution. In its $(a,A)$ vector-matrix representation the initial probability vector is $a = \begin{bmatrix} a^{(\text{BPH})} & a^{(\text{HE})} \end{bmatrix}$ where 
\[
a^{(\text{BPH})}=\left[\bar{G}(\log n)~~~\bar{G}(\log (n/2))\!-\!\bar{G}(\log n)~~...~~\bar{G}(0)\!-\!\bar{G}(\log (n/(n-1)))\right]
\] 
and $a^{(\text{HE})}=\left[p_1~~p_2~~...~~p_k\right]$. Normalization of $a$ is guaranteed by $\bar{G}(0)=1-\sum_{j=i}^k p_i$. The infinitesimal generator is
\[
A =
\begin{bmatrix}
A^{(\text{BPH})} & \mathbf{0} \\
\mathbf{0} & A^{(\text{HE})}
\end{bmatrix}
\]
where \(\mathbf{0}\) denotes zero matrices of appropriate size, indicating that there are no transitions from states of the HE component to states of the BPH component and vice versa.
\(A^{(\text{BPH})} \in \mathbb{R}^{n \times n}\) is a bidiagonal matrix corresponding to the BPH component, and \(A^{(\text{HE})} \in \mathbb{R}^{k \times k}\) is a diagonal matrix for the HE component. These are defined respectively as
\[
A^{(\text{BPH})} =
\begin{bmatrix}
-1 & 1 & 0 & \cdots & 0 \\
0 & -2 & 2 & \cdots & 0 \\
\vdots & \ddots & \ddots & \ddots & \vdots \\
0 & \cdots & 0 & -(n-1) & n-1 \\
0 & \cdots & \cdots & 0 & -n
\end{bmatrix}
\quad\quad
A^{(\text{HE})} =
\begin{bmatrix}
-\lambda_1 & 0 & \cdots & 0 \\
0 & -\lambda_2 & \cdots & 0 \\
\vdots & \vdots & \ddots & \vdots \\
0 & 0 & -\lambda_{k-1} & 0 \\
0 & 0 & \cdots & -\lambda_k
\end{bmatrix}
\]

\subsection{Optimizing data initialization for the HE tail approximation}

As already mentioned, the method proposed in \cite{[FELD98a]} requires a set of points where the HE CCDF aims to closely match the target CCDF. The selection of these points is a non-trivial task.

The main objective of this part of the study is to find the optimal set of points, which we call below parameters, to initialize the HE tail approximation of the BPH\_HE model. To achieve this, we formulate a supervised regression problem where the target values are the actual probabilities given by the CCDF we aim to approximate (denoted by $\bar{F}$), and the predictions are provided by the CCDF of the approximating BPH\_HE (denoted by $\hat{\bar{F}}$).

The two CCDFs will be compared at a number of points $z_1,z_2,...,z_n$, and, following the standard notation in supervised learning, we denote the value of $\bar{F}$ and $\hat{\bar{F}}$ at these points by:
\[
y_i=\bar{F}(z_i)~~~\mbox{and}~~~\hat{y}_i=\hat{\bar{F}}(z_i)
\]
A good approximation will generate similar $y_i$ and $\hat{y}_i$ values. The points $z_1,...,z_n$ can be selected to reflect the region of interest of the target CCDF.
For simplicity, $z_i$ are chosen to be $n$ evenly spaced points over the domain of interest. While focusing on the tail could improve precision in that region, uniform spacing ensures better stability and convergence during optimization with gradient-based methods like Adam~\cite{kingma2017adammethodstochasticoptimization}.


Two of the most applied loss functions in regression are the Mean Squared Error (MSE) and the Mean Absolute Error (MAE):
\[
\text{MSE} = \frac{1}{n} \sum_{i=1}^{n} (y_i - \hat{y}_i)^2, \quad \text{MAE} = \frac{1}{n} \sum_{i=1}^{n} |y_i - \hat{y}_i|
\]
\noindent 
Given the small differences in the tail region, using the MAE in this context is a more convenient choice.

Due to the fact that the model is probabilistic, it has a set of constraints to follow, namely, we must have $0 < p_i < 1$ and $\lambda_i > 0$. To ensure that the optimizer does not converge to solutions that violate these constraints, the original MAE loss is penalized when such constraints are not respected. The final loss function consists of three components: the Mean Absolute Error, a penalty term that enforces the constraint $0 < p_i < 1$, and an additional penalty that prevents negative values of $\lambda_i$.

Mathematically, the loss $\mathcal{L}$ is formulated as follows:   \begin{equation}
\label{eq:finalloss}
\begin{aligned}
\mathcal{L} ={} & \; w_{\text{mae}} \cdot \frac{1}{n} \sum_{i=1}^{n} |y_i - \hat{y}_i| 
+ w_{\lambda} \cdot \frac{1}{k} \sum_{i=1}^{k} \mathbf{1}(\lambda_i < 0) \\
& + w_{p} \cdot \sum_{i=1}^{k} \left[ \max(0, -p_i) + \max(0, p_i - 1) \right]
\end{aligned}
\end{equation}
\noindent
where $w_{\text{mae}}$, $w_{\lambda}$ and $w_{p}$ are the normalized weights for each loss term, $\mathbf{1}(\lambda_i \leq 0)$ is an indicator function to count the number of negative values in $\lambda$ and $\sum_{i=1}^{k} \left[ \max(0, -p_i) + \max(0, p_i - 1) \right] $ penalizes $p_i$ values that fall outside the range $[0,1]$. 

The above loss function has been optimized using an Adam optimizer, a gradient descent-like algorithm, with initialization values following a power-of-two law: $x_i=2^{i-1}$ with $i=1,2,...,2k$ where $k$ is the number of exponential components. With this initialization, the tail is covered in the interval $[1,2^{k-1}]$. If the tail region we aim to consider is different and it is given by $[x_{\min},x_{\max}]$, then we use the following initialization parameters
\[
x'_i = x_{\min} + \frac{(x_i - x_1) (x_{\max} - x_{\min})}{x_{2k} - x_{1}}~~~~i=1,2,...,2k
\]
which maps the interval $[1,2^{k-1}]$ to the interval $[x_{\min},x_{\max}]$ through a linear transformation.

\section{Experimental setup\label{sec:exp}}

\begin{figure}[H]
    \centering
    \begin{tabular}{cc}
      \includegraphics[width=0.425\textwidth]{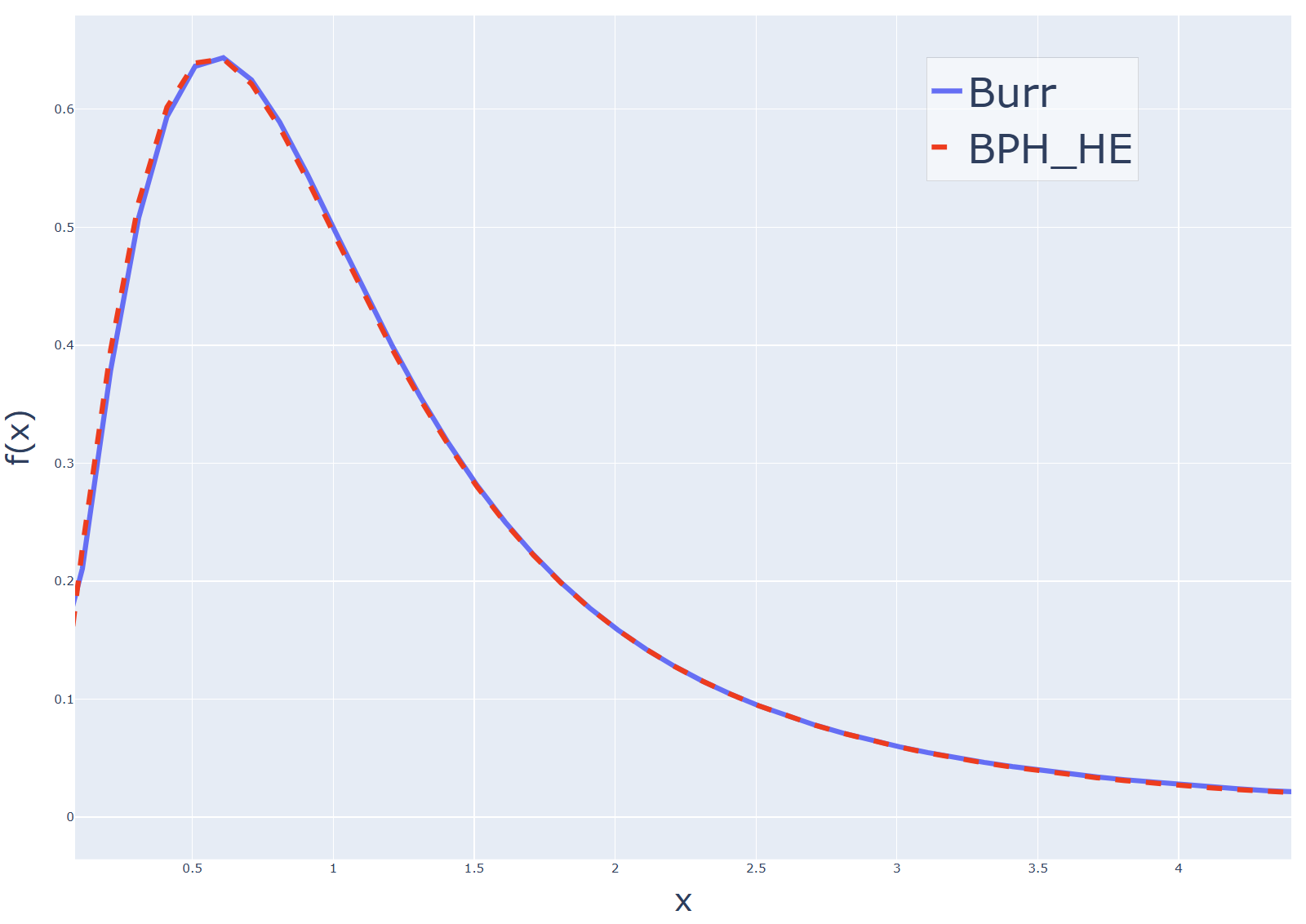}   & \includegraphics[width=0.425\textwidth]{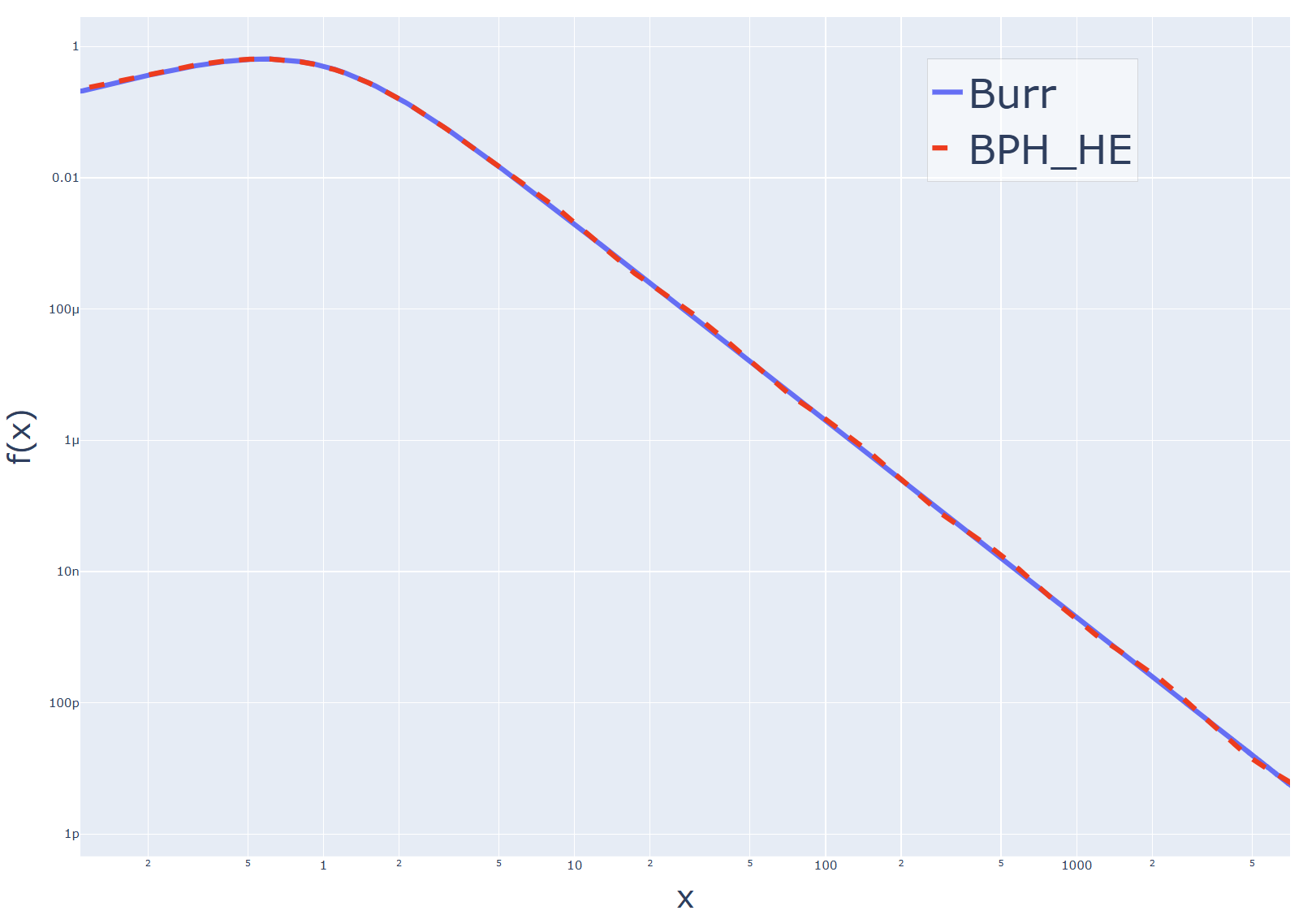} \\
      (a) body, linear plot & 
      (b) tail, log scale plot \\

      \includegraphics[width=0.425\textwidth]{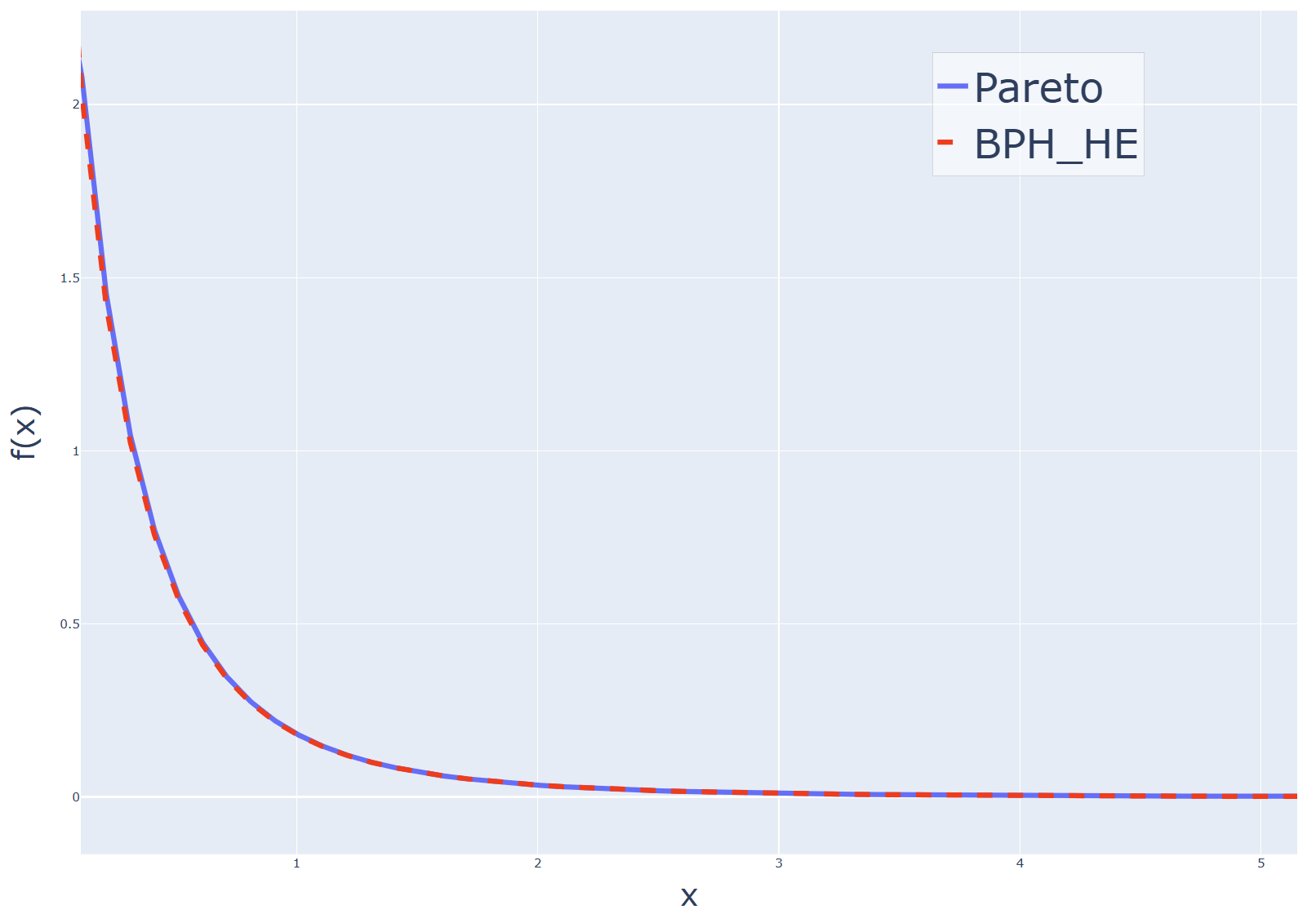}   & \includegraphics[width=0.425\textwidth]{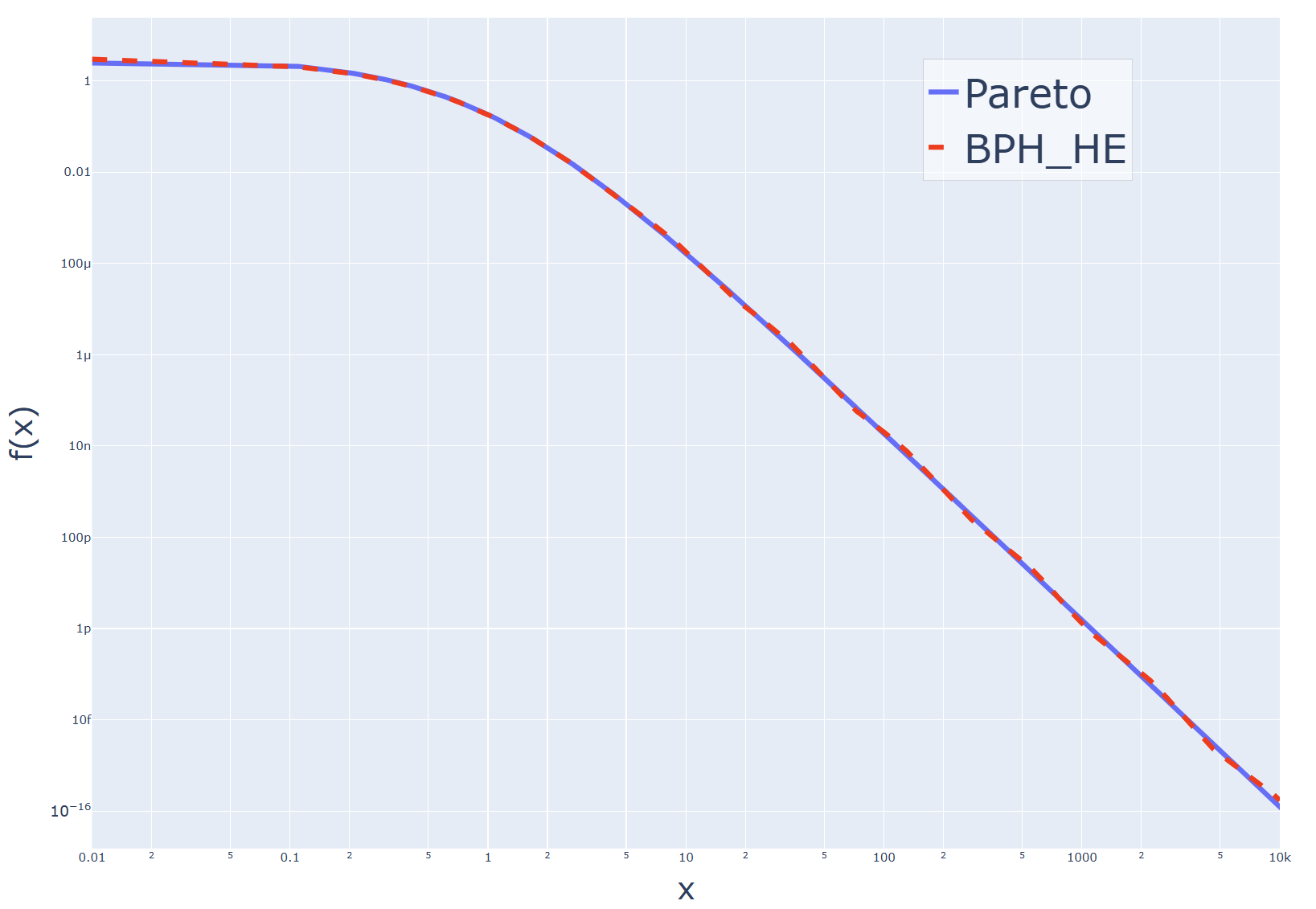} \\
      (c) body, linear plot & 
      (d) tail, log scale plot \\
      
    \end{tabular}
        \caption{PDF of BPH\_HE approximation of a Burr distribution (a, b) and a Pareto distribution (c, d). The Burr case corresponds to Figures~\ref{fig:bph_burr} and~\ref{fig:he_burr}.}
    \label{fig:grid}
\end{figure}

To assess the approximation accuracy, we run a number of experiments on well-known heavy-tailed distributions. 
\Cref{fig:grid} shows the approximation for the previously used Burr distribution and a Pareto-type distribution with CCDF 
\[
\bar{F}(x) = \frac{1}{(x+1)^{3.1}}
\]
with the corresponding BPE\_HE 
approximations.  Plots highlight a very good approximation of both the body and tail region of the distributions. 

One significant challenge in fitting probability distributions to real-world data is accurately matching statistical indices, particularly the mean and the coefficient of variation. Heavy-tailed distributions are particularly challenging because their tail behavior greatly influences these indices. If the fitted distribution does not closely match the target probability distribution, using it in real-world applications can lead to imprecise conclusions.

Table~\ref{tab:model_comparison} compares the considered models, BPH, HE, and BPH\_HE, from the perspective of capturing the mean and the coefficient of variation (CV) of a Pareto distribution. The combined model, that is, BPH\_HE distributions, clearly outperforms BPH and HE distributions. 

\setlength{\tabcolsep}{3pt}
\begin{table}[t]
    \centering
    \renewcommand{\arraystretch}{0.9}
    \begin{tabular}{lccccc}
        \toprule\begin{tabular}{c}
           \textbf{Model}
        \end{tabular} & \begin{tabular}{c} \textbf{MAE} \end{tabular} & \begin{tabular}{c}\textbf{Mean} \\ \textbf{ (Real)} \end{tabular} & \begin{tabular}{c}\textbf{Mean} \\ \textbf{(Approx)} \end{tabular}& \begin{tabular}{c}\textbf{Coef of Var}\\ \textbf{ (Real)} \end{tabular} & \begin{tabular}{c} \textbf{Coef of Var}\\ \textbf{(Approx)} \end{tabular} \\ 
        \midrule
        BPH      & 0.000019  & 0.52877  & 0.525738  & 1.444789  & 1.170760  \\
        HE    & 0.000014  & 0.52877  & 0.541833  & 1.444789  & 1.416929  \\
        BPH\_HE & \textbf{0.000001}  & 0.52877  & \textbf{0.528773}  & 1.444789   & \textbf{1.449654}   \\
        \bottomrule
    \end{tabular}
    \vspace{0.5 em}
    \caption{Statistical indices across various models on a Pareto distribution.}
    \label{tab:model_comparison}
\end{table}
Table \ref{tab:distribution_comparison} presents results obtained from fitting the BPH\_HE model to previously described Burr and Pareto, and other heavy-tailed cases such as the Weibull and Lognormal, whose CCDF are given by, respectively:
\[
\bar{F}(x) = e^{-\left(\tfrac{x}{5}\right)^{0.2}}
\]
and
\[
\bar{F}(x) = 1 - \Phi\!\left(\frac{\ln(x) - 1}{2}\right)
\]
respectively, where $\phi(\cdot)$ denotes the CDF of the standard normal distribution. 

\setlength{\tabcolsep}{3pt} %
\begin{table}[b]
    \centering
    \renewcommand{\arraystretch}{0.9}
    \begin{tabular}{lccccc}
        \toprule
        \begin{tabular}{c}
           \textbf{Distribution}
        \end{tabular} & \begin{tabular}{c} \textbf{MAE} \end{tabular} & \begin{tabular}{c}\textbf{Mean} \\ \textbf{ (Real)} \end{tabular} & \begin{tabular}{c}\textbf{Mean} \\ \textbf{(Approx)} \end{tabular}& \begin{tabular}{c}\textbf{Coef. of Var}\\ \textbf{ (Real)} \end{tabular} & \begin{tabular}{c} \textbf{Coef of Var}\\ \textbf{(Approx)} \end{tabular} \\
        \midrule
        Burr      & 0.000012  & 1.619801  & 1.620137  & 2.065175  & 2.087166  \\
        Pareto    & 0.000001  & 0.528770  & 0.528773  & 1.444789  & 1.449654  \\
        Lognormal & 0.000193  & 18.288933  & 18.473724  & 3.663488   & 3.685818   \\
        Weibull & 0.001035 & 99.583679 & 100.61869 & 2.561514 & 2.555888 \\
        \bottomrule
    \end{tabular}
    \vspace{0.5 em}
    \caption{Comparison of statistical metrics for different distributions.}
    \label{tab:distribution_comparison}
\end{table}

The Mean Absolute Error remains extremely small, showing that the approximation is highly accurate. In case of the Pareto distribution, the mean and the coefficient of variation are reproduced with near-perfect precision, while for more complex cases like the Lognormal or Weibull distributions, the approximations remain remarkably close. These results demonstrate the robustness and flexibility of the BPH\_HE approach, ensuring accurate modeling of both the body and the tail of heavy-tailed distributions.

\section{Queuing system experiment\label{sec:queue}}

To validate our method for approximating heavy-tailed distributions using the BPH\_HE model, we apply it in a queuing scenario. Many queuing systems, such as in network analysis, exhibit heavy-tailed service times due to occasional large tasks that cause high variability. These distributions are mathematically complex, making direct analysis difficult. By replacing the original heavy-tailed service time in an M/G/1 queue with its PH approximation (yielding an M/PH/1 queue), we obtain a tractable alternative. 

Validation involves comparing key performance metrics like utilization $\rho$, mean service time $E[S]$, mean waiting time $E[W]$, mean sojourn time $E[T]$, number of jobs in the system $E[N]$, and number of jobs in the queue $E[N_q]$ across the considered models: BPH, HE and BPH\_HE. The results are presented in Table~\ref{tab:queue_metrics}, where we can observe that the BPH\_HE distribution outperforms the BPH and the HE models. The final row (BPH\_HE adj.) shows results after adjusting the generator matrix of the PH model to match exactly the mean service time $m_1$ of the original heavy-tailed distribution. The adjusted matrix is computed as:
\[
A_{\text{scaled}} = \frac{\hat{m}_1}{m_1} A
\]
where $\hat{m}_1$ is the mean associated with the PH distribution with generator $A$. This adjustment makes the approximation of the utilization exact (because it depends only on the mean service time), but worsens some other measures.

\begin{table}[t]
    \centering
    \renewcommand{\arraystretch}{0.9}
    \begin{tabular}{lcccccc}
        \toprule
        \textbf{Service time} & \textbf{ $\rho$} & \textbf{ $E[S]$} & \textbf{$E[W]$} & \textbf{ $E[T]$} & \textbf{ $E[N]$} & \textbf{ $E[N_q]$} \\
        \midrule

        Pareto  & 0.238095 & 0.476190 & 0.284091 & 0.760281 & 0.380141 & 0.142045 \\
        BPH    & 0.230504 & 0.461009 & 0.213801 & 0.674809 & 0.337405 & 0.106900 \\
        HE    & 0.126288 & 0.252576 & 0.211307 & 0.463884 & 0.231942 & 0.105654 \\
         BPH\_HE & 0.246392 & 0.492785 & 0.268821 & 0.761607 & 0.380803 & 0.134410 \\ 
        BPH\_HE adj. & {0.238095} & {0.476190} & {0.248287} & {0.724477} & {0.362239} & {0.124143} \\
                \bottomrule
     \end{tabular}
     \vspace{0.5 em }
    \caption{Comparison of queues using different service time approximations.} \label{tab:queue_metrics}
\end{table}


\begin{figure}[b]
    \centering
    \begin{tabular}{cp{1cm}c}
        \includegraphics[width=0.4\textwidth]{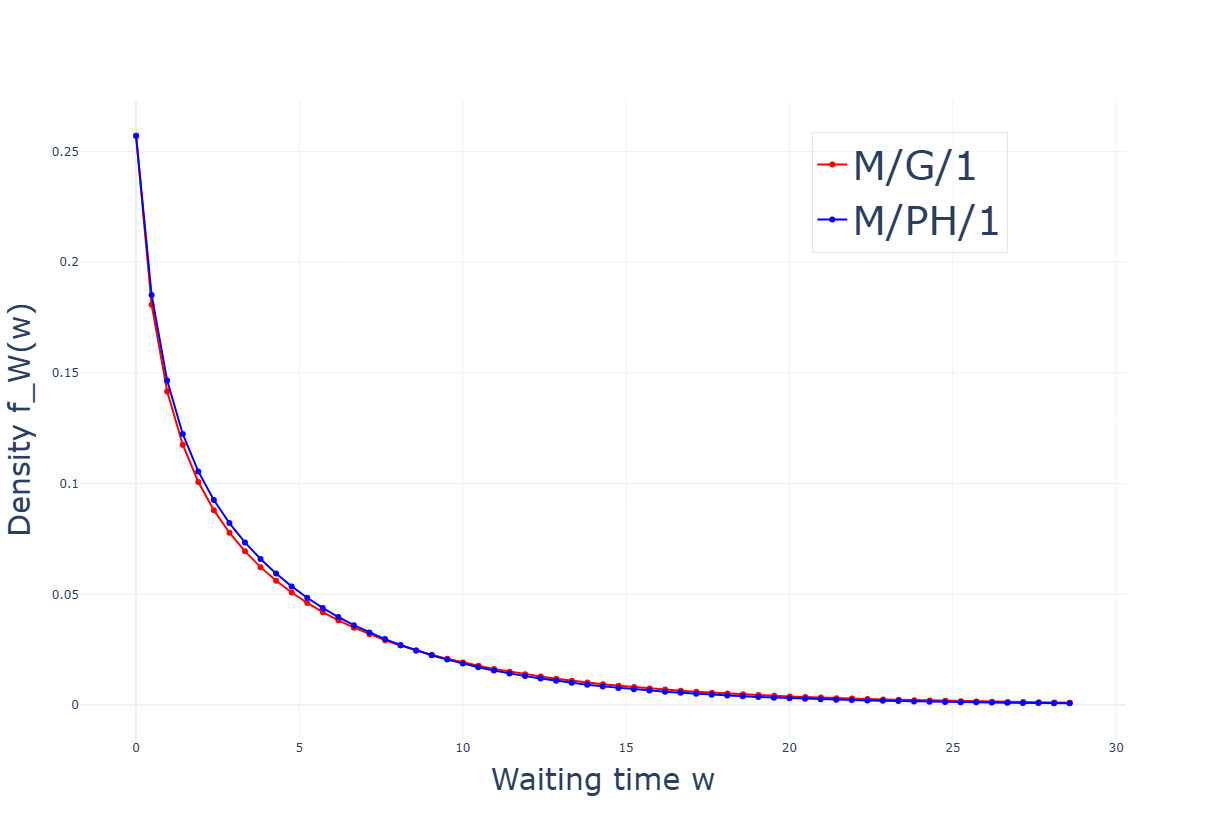}  &
        &
        \includegraphics[width=0.4\textwidth]{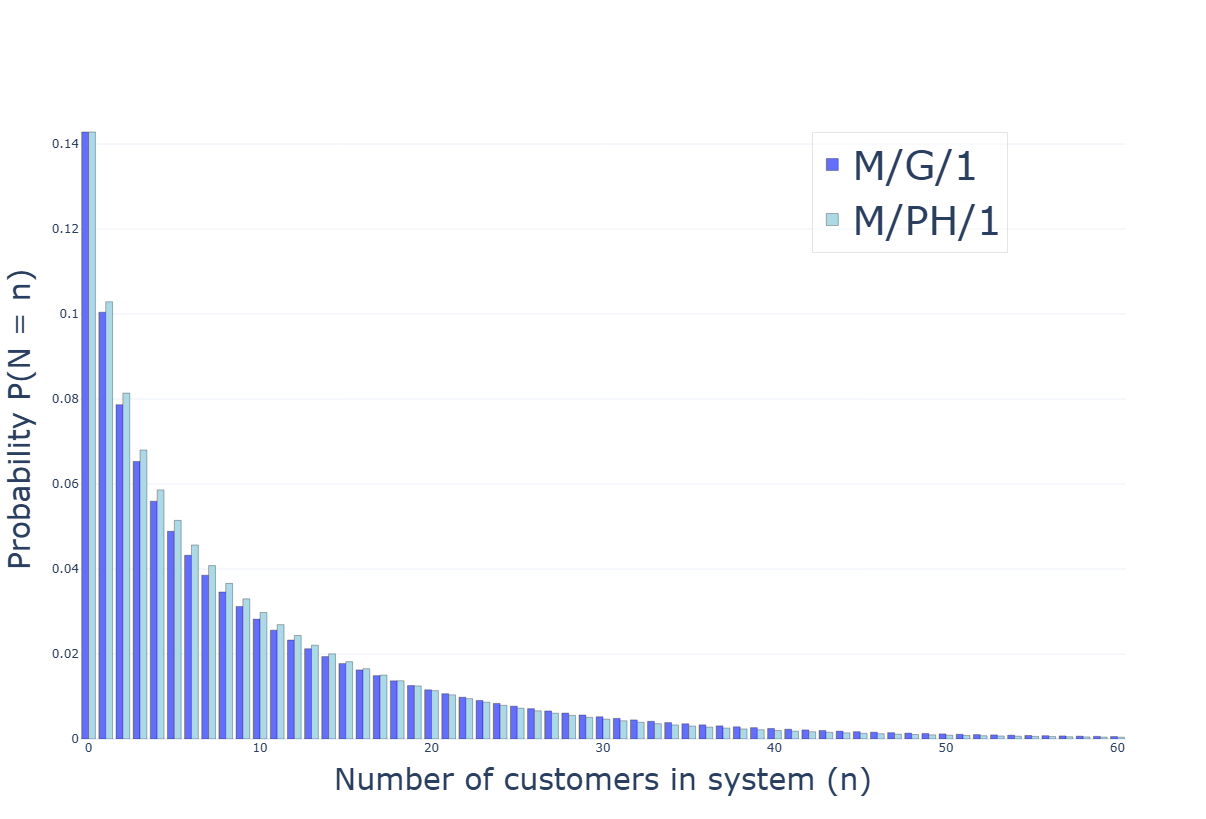} \\
        (a) Waiting times distribution & &
        (b) Queue length distribution
    \end{tabular}
        \caption{Distribution-based comparison.}
    \label{fig:queue_distribution_results}
\end{figure}

\Cref{fig:queue_distribution_results} illustrates some distribution-based analysis, comparing the waiting time (left) and queue length (right) distributions of the original (M/Pareto/1) queue and the approximate (M/PH/1) queue with BPH\_HE service time. These visualizations offer insights into the PH approximation for capturing rare but impactful events, long waiting times, and extended queue lengths.

Overall, these results validate the practical utility of approximating service times using a mixture of BPH and HE models, particularly for capturing performance in typical queuing scenarios.

\section{Conclusion and future work\label{sec:conc}}

This research proposed the BPH\_HE mixture model to approximate heavy-tailed distributions, addressing the limitations of existing methods by combining the tractability of Bernstein PH distributions with the tail fitting flexibility of hyperexponential models. For the parametrization of the hyperexponential component, a novel initialization method using gradient descent was introduced. The resulting approach accurately captures both body and tail regions of distributions like Burr, Pareto, and Lognormal. Validation through queuing theory confirmed the model’s effectiveness in applications. Future work involves applying this heavy-tail aware model in AI algorithms where rare and extreme events play a critical role, such as in generative models or risk-sensitive applications in finance and cybersecurity. With that, we hope to achieve more robust and reliable results compared to traditional Gaussian-based approaches.

\bibliographystyle{elsarticle-num}
\bibliography{main}

\begin{thebibliography}{10}
\expandafter\ifx\csname url\endcsname\relax
  \def\url#1{\texttt{#1}}\fi
\expandafter\ifx\csname urlprefix\endcsname\relax\def\urlprefix{URL }\fi
\expandafter\ifx\csname href\endcsname\relax
  \def\href#1#2{#2} \def\path#1{#1}\fi

\bibitem{foss2011introduction}
S.~Foss, D.~Korshunov, S.~Zachary, et~al., An introduction to heavy-tailed and subexponential distributions, Vol.~6, Springer, 2011.

\bibitem{stewart1994markov}
W.~J. Stewart, Introduction to the Numerical Solution of Markov Chains, Princeton University Press, Princeton, NJ, 1994.

\bibitem{HorvathV23}
A.~Horv{\'{a}}th, E.~Vicario, Construction of phase type distributions by {B}ernstein exponentials, in: Proceedings of {ASMTA} 2023, Vol. 14231 of Lecture Notes in Computer Science, Springer, 2023, pp. 201--215.

\bibitem{HORVATH2025102480}
A.~Horváth, I.~Horváth, M.~Paolieri, M.~Telek, E.~Vicario, Approximation of cumulative distribution functions by bernstein phase-type distributions, Performance Evaluation 168 (2025).
\newblock \href {https://doi.org/https://doi.org/10.1016/j.peva.2025.102480} {\path{doi:https://doi.org/10.1016/j.peva.2025.102480}}.

\bibitem{[FELD98a]}
A.~Feldman, W.~Whitt, Fitting mixtures of exponentials to long-tail distributions to analyze network performance models, Performance Evaluation 31 (1998) 245--279.

\bibitem{HoTe00}
A.~Horv\'ath, M.~Telek, Approximating heavy tailed behavior with phase-type distributions, in: Proc. of 3rd International Conference on Matrix-Analytic Methods in Stochastic models, Leuven, Belgium, 2000, pp. 191--214.

\bibitem{[NEUT75]}
M.~Neuts, Probability distributions of phase type, in: Liber Amicorum Prof. Emeritus H. Florin, University of Louvain, 1975, pp. 173--206.

\bibitem{johnson1989matching}
M.~A. Johnson, M.~R. Taaffe, Matching moments to phase distributions: Mixtures of {E}rlang distributions of common order, Stochastic Models 5~(4) (1989) 711--743.

\bibitem{bobbio2005matching}
A.~Bobbio, A.~Horv{\'a}th, M.~Telek, Matching three moments with minimal acyclic phase type distributions, Stochastic Models 21~(2-3) (2005) 303--326.

\bibitem{horvath2007matching}
A.~Horv{\'a}th, M.~Telek, Matching more than three moments with acyclic phase type distributions, Stochastic Models 23~(2) (2007) 167--194.
\newblock \href {https://doi.org/10.1080/15326340701300712} {\path{doi:10.1080/15326340701300712}}.

\bibitem{BuKr09}
P.~Buchholz, J.~Kriege, A heuristic approach for fitting maps to moments and joint moments, in: Proceedings of the 2009 Sixth International Conference on the Quantitative Evaluation of Systems, QEST '09, IEEE Computer Society, USA, 2009, p. 53–62.
\newblock \href {https://doi.org/10.1109/QEST.2009.36} {\path{doi:10.1109/QEST.2009.36}}.

\bibitem{sherzer2025unconstrainedoptimizationapproachmoment}
E.~Sherzer, Y.~Resheff, M.~Telek, An unconstrained optimization approach to moment fitting with phase type distributions (2025).
\newblock \href {http://arxiv.org/abs/2505.20379} {\path{arXiv:2505.20379}}.

\bibitem{horvath2024phase}
A.~Horv{\'a}th, M.~Telek, Phase Type Distributions: Theory and Application, Wiley-ISTE, 2024.

\bibitem{asmussen1996fitting}
S.~Asmussen, O.~Nerman, M.~Olsson, Fitting phase-type distributions via the {EM} algorithm, Scandinavian Journal of Statistics 23~(4) (1996) 419--441.

\bibitem{okamura2011refined}
H.~Okamura, T.~Dohi, K.~S. Trivedi, A refined {EM} algorithm for {PH} distributions, Performance Evaluation 68~(10) (2011) 938--954.

\bibitem{[BOBB94a]}
A.~Bobbio, M.~Telek, A benchmark for {PH} estimation algorithms: results for {A}cyclic-{PH}, Stochastic Models 10 (1994) 661--677.

\bibitem{Haddad1997}
S.~Haddad, P.~Moreaux, G.~Chiola, Efficient handling of phase-type distributions in generalized stochastic petri nets, in: P.~Azéma, G.~Balbo (Eds.), Application and Theory of Petri Nets 1997, Vol. 1248 of Lecture Notes in Computer Science, Springer, 1997, pp. 175--194.
\newblock \href {https://doi.org/10.1007/3-540-63139-9\_36} {\path{doi:10.1007/3-540-63139-9\_36}}.

\bibitem{neuts1981matrix}
M.~F. Neuts, Matrix-Geometric Solutions in Stochastic Models: An Algorithmic Approach, Johns Hopkins University Press, Baltimore, 1981, foundational work on matrix-geometric methods for solving stochastic models with phase-type distributions.

\bibitem{latouche1999introduction}
G.~Latouche, V.~Ramaswami, Introduction to Matrix Analytic Methods in Stochastic Modeling, ASA-SIAM, Philadelphia, 1999, comprehensive treatment of matrix-analytic methods for phase-type models.

\bibitem{Cu85}
A.~Cumani, Esp - {A} package for the evaluation of stochastic {P}etri nets with phase-type distributed transition times, in: International Workshop on Timed Petri Nets, IEEE Computer Society, USA, 1985, p. 144–151.

\bibitem{Horvath2002PhFit}
A.~Horváth, M.~Telek, Phfit: A general phase-type fitting tool, in: Computer Performance Evaluation: Modelling Techniques and Tools (TOOLS 2002), Vol. 2324 of Lecture Notes in Computer Science, Springer, 2002, pp. 82--91.
\newblock \href {https://doi.org/10.1007/3-540-46029-2_5} {\path{doi:10.1007/3-540-46029-2_5}}.

\bibitem{HyperStarTool}
Hyperstar: A tool for fitting hyper-erlang distributions, \url{https://www.mi.fu-berlin.de/inf/groups/ag-dds/Tools/HyperStar/index.html}, {FU} Berlin, {AG DDS Group}, Accessed: 2025-08-28.

\bibitem{Horvath2017BuTools}
G.~Horváth, M.~Telek, Butools 2: A rich toolbox for markovian performance evaluation, in: Proceedings of the 10th EAI International Conference on Performance Evaluation Methodologies and Tools, ACM, 2017.
\newblock \href {https://doi.org/10.4108/eai.25-10-2016.2266400} {\path{doi:10.4108/eai.25-10-2016.2266400}}.

\bibitem{AFOLALU20212884}
S.~Afolalu, O.~Ikumapayi, A.~Abdulkareem, M.~Emetere, O.~Adejumo, A short review on queuing theory as a deterministic tool in sustainable telecommunication system, Materials Today: Proceedings 44 (2021) 2884--2888, international Conference on Materials, Processing and Characterization.
\newblock \href {https://doi.org/https://doi.org/10.1016/j.matpr.2021.01.092} {\path{doi:https://doi.org/10.1016/j.matpr.2021.01.092}}.

\bibitem{HarcholBalter2021}
M.~Harchol-Balter, Open problems in queueing theory inspired by datacenter computing, Queueing Systems 97~(1) (2021) 3--37.
\newblock \href {https://doi.org/10.1007/s11134-020-09684-6} {\path{doi:10.1007/s11134-020-09684-6}}.

\bibitem{riskfinanceht}
G.~Zi-Yi, Heavy-tailed distributions and risk management of equity market tail events, Journal of Risk and Control 4~(1) (2017).

\bibitem{Piantadosi2014}
S.~T. Piantadosi, Zipf’s word frequency law in natural language: A critical review and future directions, Psychonomic Bulletin \& Review 21~(5) (2014) 1112--1130.
\newblock \href {https://doi.org/https://doi.org/10.3758/s13423-014-0585-6} {\path{doi:https://doi.org/10.3758/s13423-014-0585-6}}.

\bibitem{llmzifs}
S.~Su, H.~Zhang, Z.~Wang, Evaluating large language models on twitter based on hashtag dynamics and scaling properties, in: H.~Zhang, J.~Su, J.~Shang (Eds.), Intelligent Multilingual Information Processing, Springer Nature Singapore, Singapore, 2025, pp. 236--248.

\bibitem{kingma2017adammethodstochasticoptimization}
D.~P. Kingma, J.~Ba, \href{https://arxiv.org/abs/1412.6980}{Adam: A method for stochastic optimization} (2017).
\newline\urlprefix\url{https://arxiv.org/abs/1412.6980}

\end{thebibliography}

\end{document}